# Widely Tunable Photonic Filter Based on Equivalent Chirped Four-Phase-Shifted Sampled Bragg Gratings

Simeng Zhu*, Bocheng Yuan, Mohanad Al-Rubaiee, Yiming Sun, Yizhe Fan, Ahmet Seckin Hezarfen, Stephen J. Sweeney, John H. Marsh and Lianping Hou

James Watt School of Engineering, University of Glasgow, Glasgow, G12 8QQ, UK

*Supporting Information Placeholder*

**ABSTRACT:** We have developed an integrated dual-band photonic filter (PF) utilizing equivalent chirped four-phase-shifted sidewall-sampled Bragg gratings (4PS-SBG) on a silicon-on-insulator (SOI) platform. Using the reconstruction equivalent-chirp technique, we designed linearly chirped 4PS Bragg gratings with two π-phase shifts (π-PS) positioned at 1/3 and 2/3 of the grating cavity, introducing two passbands in the +1st order channel. Leveraging the significant thermo-optic effect of silicon, dual-band tuning is achieved through integrated micro-heaters (MHs) on the chip surface. By varying the injection currents from 0 to 85 mA into the MHs, the device demonstrates continuous and wide-range optical frequency division (OFD) performance, with the frequency interval between the two passbands adjustable from 52.1 GHz to 439.5 GHz. Four notable frequency division setups at 100 GHz, 200 GHz, 300 GHz, and 400 GHz were demonstrated using a 100 GHz, 1535 nm semiconductor passive mode-locked laser as the light source.

**KEYWORDS:** *sampled Bragg grating, photonic filter, tunable filter, micro heater, optical frequency division, equivalent reconstruction technology.*

## INTRODUCTION

The increasing demand for higher bandwidth is positioning photonic systems as prime candidates for future telecommunications and radar technologies.[1] Integrated photonic systems offer ultra-wideband performance within compact dimensions and can naturally interface with fiber optic networks for signal transmission.[2-4] To address the evolving operational demands of wavelength-division multiplexing (WDM) systems, data center optical networks, and microwave photonics, there is a pressing need for the development of tunable and reconfigurable multi-band photonic filters (PFs).[5] Multi-channel PFs can effectively suppress unwanted signals in congested radio frequency (RF) spectrum environments while preserving multiple useful signals.[2,6] As RF systems advance to operating frequencies above 30 GHz, into the millimeter-wave band, these filters must also provide wideband frequency tunability.[3] Furthermore, they need to be compact, lightweight, and energy-efficient to meet the stringent requirements of future RF applications.[7-8]

PFs utilize the broadband tunability and reconfigurability of optical components to overcome the frequency tuning limitations that have constrained traditional RF filtering technologies over the past 20 years.[9-12] Recent research has reported various designs for multi-frequency PFs, which can be realized by cascading multiple single-channel PFs. For instance, this can be accomplishes by employing two micro-ring resonator (MRR)-loaded Mach-Zehnder Interferometers (MZI)[13] to achieve complex rectangular spectral responses, with a 100 GHz free spectral range (FSR) limitation, or two cascaded distributed feedback Bragg grating resonators (DFBR)[14] with a -3 dB bandwidth of 24 GHz and a cavity length of 875 μm.[15] Another straightforward approach to achieving single-channel filtering is through the nonlinear effect of Stimulated Brillouin Scattering (SBS).[16] In tunable filters based on silicon photonics, SBS is favored for its high selectivity and narrow bandwidth, enabling ultra-narrowband filtering of optical signals with -3 dB bandwidths in the MHz range, providing fine spectral resolution.[17-18] However, SBS has several drawbacks. Firstly, since the local oscillator (LO) tone required for SBS must be generated by an external intensity modulator and an extra bandpass filter working together, the modulator and acoustic wave frequency limit its application in high-speed optical communication systems.[19] Moreover, the latest SBS-based filters reported in the literature have limitations, including a narrow tuning range (maximum 15 GHz), complex fabrication processes (due to the special nature of SBS waveguide materials), high waveguide transmission losses (2 dB/cm), and large on-chip area requirements (owing to the long cavity length of SBS waveguides, typically 5.5 cm).[18-21]

Another approach to achieving dual-frequency PF is to directly implement multiple passbands using a single PF. Examples include equivalent phase shift fiber Bragg gratings (EPS-FBG),[5] based on phase-modulation-to-intensity modulation (PM-IM) conversion, which feature a cavity length of 32 cm, a -3 dB bandwidth of 167 MHz, and a maximum tuning range of just 7.4 GHz. Similarly, MRR offers a -3 dB bandwidth of 15 GHz and a free spectral range (FSR) of 282 GHz,[22] but does not allow for independent tuning of each passband. And the EO-modulated phase shift Bragg grating (PS-BG),[23] which has a 14 pm linewidth dual channel with a modulation range of 113.4 GHz, but its passbands are also not independently tunable.

In this study, we designed and fabricated a tunable dual-passband PF on an SOI platform, enabling independent tuning of two passbands on a single grating waveguide. This approach utilizes an equivalent chirped[24] four-phase-shifted sidewall-sampled Bragg gratings (4PS-SBG) with two π-phase shifts (π-PSs) and the thermo-optic interaction of two micro-heaters (MHs). The equivalent chirp of the grating spatially separates the photons of the two passbands, enabling independent tuning of the passband wavelengths. This equivalent chirp is achieved by linearly modulating the micron-scale sampling period, significantly reducing manufacturing complexity compared to modulating the nanometer-scale seed grating period.[25] Two MHs are placed at the positions of the two PS points. By leveraging the thermo-optic effect to alter the phase amplitude of the two PS, the positions of the passbands can be precisely adjusted.[26] The integration of the 4PS-SBG and MH offers greater design flexibility compared to multi-MRR and SBS filters, enabling the development of narrowband, tunable, and customizable multi-passband filters. The integrated PF demonstrated a frequency spacing tuning range of 387 GHz, achieving a -3 dB bandwidth of 16 GHz for a single passband. Additionally, by using a semiconductor mode-locked laser (SMLL) in optical frequency division experiments, the PF suppressed interference signals within the stopband spectral range, achieving excellent optical frequency division performance from 100 GHz to 400 GHz, with a maximum side-mode suppression ratio (SMSR) exceeding 10 dB.

Compared to conventional MZI and MRR-based filters, the proposed PF provides narrower passband frequencies and a wide tunable frequency range that is not constrained by the FSR.[2] Additionally, our approach reduces the number of resonators needed to process multiple signals, offering a compact solution for photonic integrated circuits (PICs). The devices were fabricated on a standard SOI platform using established manufacturing techniques and waveguide geometries, facilitating wafer-level integration of other multi-channel filter components on the same platform.

## DESIGN AND METHOD

For the refractive index modulation of the sampled grating, it can be written as the following:[27]

$$\Delta n(z) = s(z) \cdot \frac{1}{2} \cdot (\Delta n_0 e^{-i\frac{2\pi z}{\Lambda_0}} + c.c) \quad (1)$$

where $\Delta n_0$ is the refractive index modulation amplitude of the seed grating, $z$ is the position in the grating cavity, and $\Lambda_0$ is the seed grating period. $s(z)$ is the sampling function. The Fourier expansion of the $s(z)$ can be expressed as follows:

$$s(z) = \sum_m S_m e^{-i\frac{2m\pi z}{P_s}} \quad (2)$$

where $m$ represents the Fourier series, $S_m$ denotes the corresponding channel intensity coefficient, and $P_s$ refers to the sampling period. For the conventional sampled Bragg grating (C-SBG), each section, including the grating and non-grating parts, has a length equal to half of the sampling period $P_s$. Within a complete sampling period $P_s$, the corresponding $s(z)$ for the C-SBG is given as follows:[28]

$$s(z)_{C-SBG} = \begin{cases} 1, & 0 < z \leq \frac{P_s}{2} \\ 0, & \frac{P_s}{2} < z \leq P_s \end{cases} \quad (3)$$

Here, +1st channel ($m = +1$) is used as the filter's working channel. By calculating the Fourier coefficients, the coupling coefficient $\kappa$ of the +1st sub-grating in the C-SBG is $1/\pi$ times that of the uniform Bragg grating (UBG). It is important to note that a lower $\kappa$ results in a narrower stopband width and a weaker extinction ratio (ER), which for filters, translates to a narrower tuning range and a lower SMSR. The current mainstream solution to address the low ER of C-SBG is to compensate for the insufficient $\kappa$ by designing a longer cavity length. However, in SOI-based single-mode waveguides, increasing the cavity length results in more scattering, which leads to higher propagation losses. Additionally, this approach is not conducive to the miniaturization and integration of photonic chips. To address the issues associated with C-SBG, we implemented a 4PS-SBG structure in our design. In comparison, the 4PS-SBG structure divides each $P_s$ into four equal segments, with each adjacent segment having a $\pi/2$ phase shift. The new corresponding $s(z)$ is as follows:[28]

$$s(z)_{4PS-SBG} = \begin{cases} 1, & 0 < z \leq \frac{P_s}{4} \\ e^{i\frac{\pi}{2}}, & \frac{P_s}{4} < z \leq \frac{P_s}{2} \\ e^{i\pi}, & \frac{P_s}{2} < z \leq \frac{3P_s}{4} \\ e^{i\frac{3\pi}{2}}, & \frac{3P_s}{4} < z \leq P_s \end{cases} \quad (4)$$

Similarly, by calculating Fourier coefficients, we found that the 4PS-SBG structure results in a $\kappa$ value for its +1st sub-grating channel that is approximately 0.9 times that of a uniform Bragg grating (UBG). This means that when using the +1st passband as the working channel for the filter, the 4PS-SBG structure allows for a shorter cavity length compared to the C-SBG structure, while achieving the same ER performance and a wider stopband. Figure 1(a) and 1(b) respectively show the schematic and the transmission spectra of UBG, C-SBG, and 4PS-SBG, calculated using the Transfer Matrix Method (TMM).[29,30] Notably, the 4PS sampling structure not only enhances the +1st sub-grating channel by approximately three times but also effectively suppresses the 0th order channel, preventing crosstalk from spurious signals in the 0th order channel to the microwave signal, compared to C-SBG.

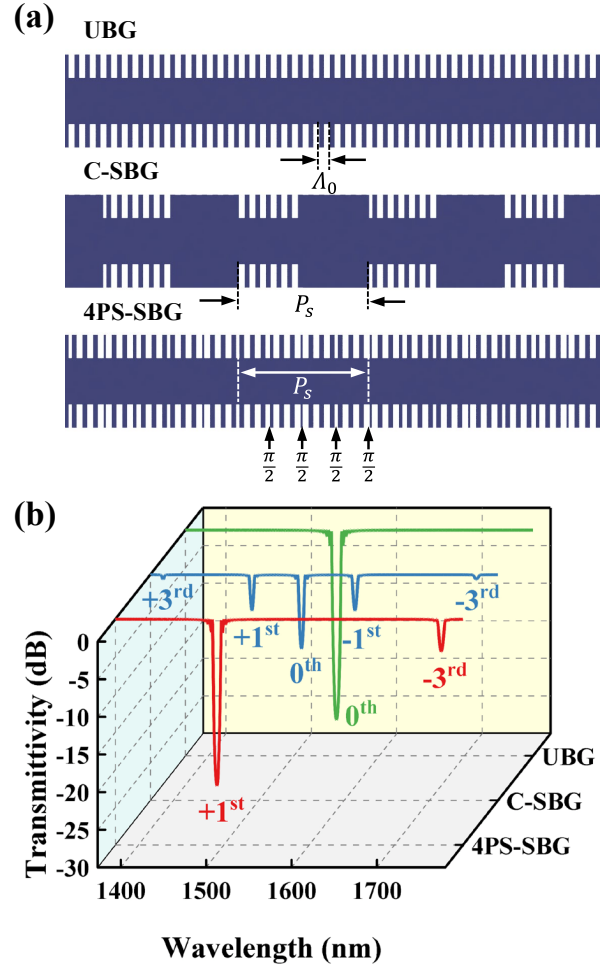

**Figure 1.** (a) Schematic and (b) transmission spectrum of the UBG, C-SBG and 4PS-SBG.

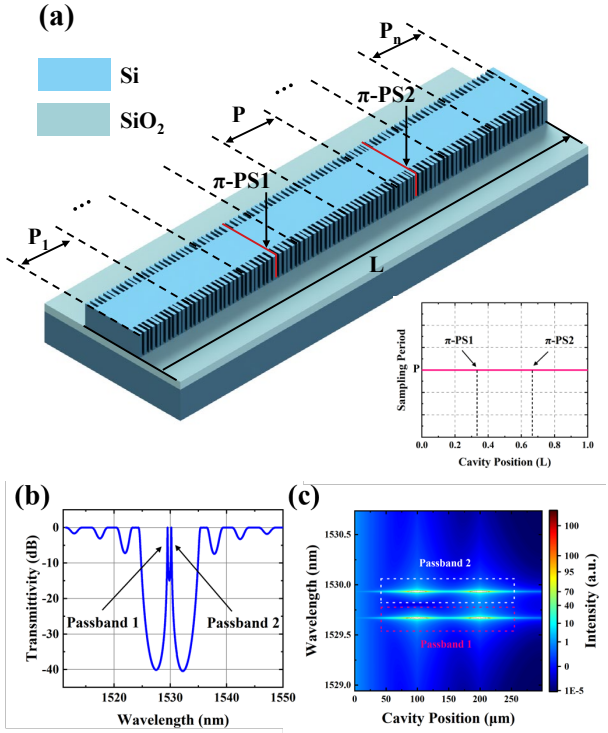

**Figure 2.** (a) Schematic representation of the uniform 4PS-SBG and sampling period distribution. (b) Transmission spectrum of the uniform 4PS-SBG. (c) Photon distribution along the cavity of the uniform 4PS-SBG.

Typically, introducing two π-phase shift points in a C-SBG-based cavity can generate two transmission peaks in all sub-channels. Figure 2(a) illustrates a schematic of a uniform 4PS-SBG with two π-PSs. To generate two distinct passbands, two π phase shifters (π-PS1 and π-PS2) are strategically placed at one-third and two-thirds of the total cavity length ($L$). Here, $L$ is 300 μm, the seed grating period $\Lambda_0$ is 351 nm, and the sampling period $P_S$ is 3.113 μm, which places the +1st channel at 1530 nm. The transmission spectrum and photon distribution of the entire grating are crucial for predicting the filter's performance. Therefore, we used the TMM to obtain the transmission spectrum and the corresponding photon distribution of the uniform 4PS-SBG, as shown in Figure 2(b). In the transmission stopband, two distinct passbands (referred to as passband 1 and passband 2) can be clearly observed, which are introduced by PS1 and PS2. Similarly, the photon distribution can also be calculated using the TMM. Figure 2(c) shows the distribution of photons along the cavity length within the stopband wavelength range. The photon distributions of passband 1 and passband 2 nearly overlap, indicating that the formation of these passbands results from the combined accumulation of incident light at the two PS points, with the light then transmitted through the other end of the grating. This overlap causes the optical fields of both passbands to change simultaneously when the phase shift magnitude at either PS point is adjusted, making independent tuning of the passbands impossible. This behavior arises from the distributed feedback nature of the uniform grating, where no fixed reflection points exist at specific locations.

To address the aforementioned issue, we introduce an equivalent chirped grating design (Figure 3(a)), which ensures that light of a specific frequency is reflected by the specific sections of the chirped grating.[31] By introducing a linear chirp to the sampled grating, the sampling periods at the two PS locations differ, leading the two passbands to originate from different optical cavities, which enables independent tuning of the passbands. The linear chirp sampling period distribution satisfies the following equation:

$$P_n = P_1 + C \cdot z_n \tag{5}$$

where $P_1$ is the first sampling period, $P_n$ represents the n$^{th}$ sampling period at start position of $z_n$. $C$ refers to the chirp rate, defined as the ratio of the difference between the maximum and minimum sampling periods ($P_n - P_1$) to the cavity length $L$. In our device, the chirp rate $C$ is set to 200 nm/mm. Due to the introduction of the chirped structure, the sampling periods corresponding to the two π-phase shifts are different. Each phase shift introduces a transmission peak at a wavelength that satisfies the local Bragg condition. Figure 3(b) is the transmission spectrum of the chirped 4PS-SBG calculated using the TMM. Two narrow passbands, referred to as passband 1 and passband 2 (respective center wavelengths are $\lambda_1$ and $\lambda_2$) are clearly observed within the transmission stopband, aligning with the phase discontinuities caused by PS1 and PS2, respectively. The spatial photon distribution of the two passbands is shown in Figure 3(c). Compared to uniform sampled gratings, the optical fields of passband 1 and passband 2 are spatially concentrated around π-PS1 and π-PS2, respectively, with their photon distributions clearly separated. This separation indicates the mutual independence of the two passbands. In the stopband, longer wavelengths are reflected nearer the end of the grating, where the grating period is longer, while shorter wavelengths are reflected nearer the beginning of the grating, where the period is shorter.

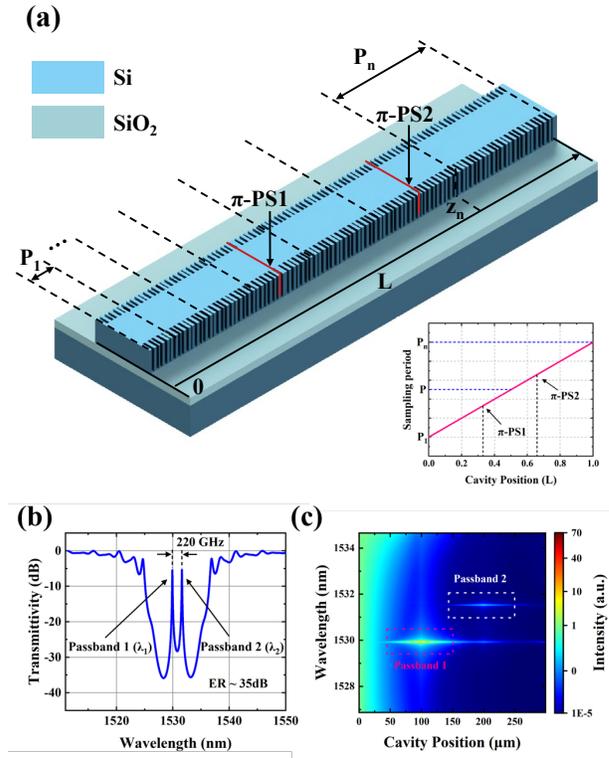

**Figure 3.** (a) Schematic representation of the chirped 4PS-SBG and sampling period distribution. (b) Transmission spectrum of the chirped 4PS-SBG. (c) Photon distribution along the cavity of the chirped 4PS-SBG.

Therefore, when we modulate the value of a specific PS, only one corresponding wavelength is adjusted, while the other remains unchanged. In Figure 4(a), we calculated the frequency spacing as a function of the values of PS1, with PS2 fixed at π, under various chirp rates. The results show that increasing only the phase shift of PS1 leads to a redshift of $\lambda_1$, while the corresponding $\lambda_2$ of PS2 remains constant, leading to a decrease in the frequency spacing. Similarly, in Figure 4(b), when only the phase shift of PS2 is increased, $\lambda_1$ remains unchanged while $\lambda_2$ is shifted away from $\lambda_1$, increasing the frequency spacing. This method achieves modulation of the filter passband spacing. A higher chirp rate results in a wider separation between the dual passbands. Our calculations show that, at a chirp rate of 200 nm/mm, the dual passband separation frequency can be tuned by over 600 GHz, marking a significant improvement compared to previous similar works.[5, 14-15]

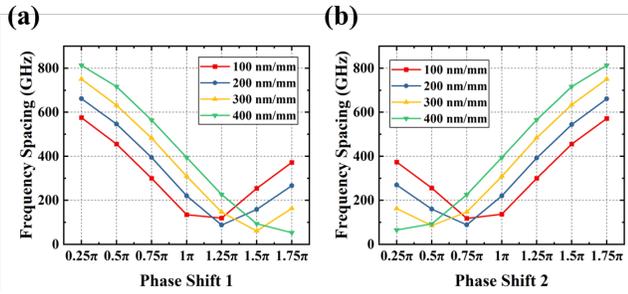

**Figure 4.** Calculated frequency spacings at different chirp rates as a function of the amplitude of PS1 (a) and PS2 (b), respectively.

Typically, thermo-optic phase shifts can be achieved either by integrating MHs on top of the waveguide or by utilizing the resistive effect of doped waveguides.[32] The former approach involves using a sufficiently thick protective cladding to prevent the metal from affecting the optical propagation modes, thereby minimizing additional optical losses caused by evanescent field. The latter method, however, provides faster response times. Despite this, the integration of metal MHs avoids the need for ion implantation and annealing, simplifying the fabrication process. Consequently, this paper uses metal MHs to achieve thermo-optic phase shifts. The silicon waveguide has a cross-section of 0.5 μm × 0.22 μm, with a buried oxide layer of 2 μm and a top cladding oxide thickness of 1 μm. The heater metal consists of titanium (Ti)/platinum (Pt)/Gold (Au) with a total thickness of 0.23 μm. The width of the heating wire is 2 μm, where Ti serves as an adhesion layer, Pt acts as the primary heating element, and gold provides protection against oxidation and corrosion.

The heating distribution of the chirped 4PS-SBG PF is simulated using COMSOL Multiphysics in a three-dimensional model (Figure 5(a)). The relationship between the phase shift of the waveguide and the temperature change, $\Delta T$, is given by:[33]

$$\Delta \phi = \frac{2\pi}{\lambda} L \left( \frac{\partial n_{eff}}{\partial T} \right) \Delta T \qquad (6)$$

Here, $n_{eff}$ represents the effective refractive index, and $L$ is the resonance cavity length. MH1 and MH2 are thin-film heaters placed on top of the cladding oxide, with a 100 μm spacing between the two heating wires, both sharing a common ground plate. In this simulation setup, we apply the electric potential only to MH2 to observe the effects of thermal crosstalk. As shown in Figure 5(a), MH1 exhibits no significant surface temperature change, while MH2 demonstrates highly localized heating. Assuming ideal heat transfer at the phase-shift point, the cross-sectional thermal distributions of MH1 and MH2 also confirm that when only one MH is working, there is no temperature change inside the silicon waveguide below the other MH, while the heat generated by the working MH penetrates highly vertically through the top cladding $SiO_2$ layer to reach the silicon waveguide below it, with little or no horizontal diffusion. Figure 5(b) shows how the phase shift amplitude of the waveguide changes under the two MHs as a function of the tuning current applied to MH2. For a tuning current of 91 mA, which induces a phase change of $1\pi$ in PS2, the phase change in PS1 under the same conditions is only $0.012\pi$. If thermal crosstalk is defined as the ratio of the phase change in the unmodulated waveguide to that in the modulated waveguide,[34] then for this device with the given MH geometric design, the thermal crosstalk is only 1.2%. This indicates that the MH distribution in this device provides excellent thermal isolation, with the thermal effects on adjacent phase shifters beyond 100 μm being negligible.

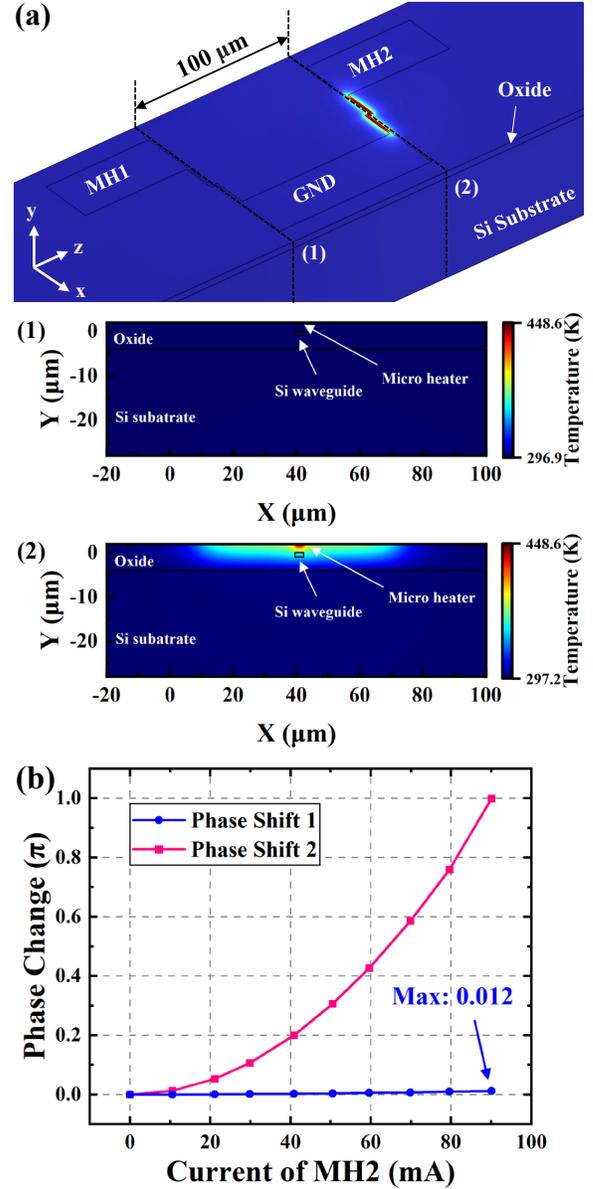

**Figure 5.** (a) Simulated temperature distribution when only MH2 is operating and the corresponding cross sections at positions (1) and (2). (b) Phase change of MH1 and MH2 when injection current is applied only to MH2.

## FABRICATION AND CHARACTERISATION

Figure 6(a) is a schematic of the proposed device. Where the 4PS-SBG section ($L = 670\ \mu m$) is at the center of the device and the two tapers ($L_{taper} = 250\ \mu m$) for connecting the 4PS-SBG to the grating couplers (GC, 12μm wide and 16 μm long) are symmetrically integrated on both sides. Buffer waveguides, each 10 μm long ($L_{buffer}$), are placed on both sides to connect the taper and the GC with varying waveguide width, ensuring a smooth optical mode transition within the waveguide. The SOI wafer has a 220 nm top silicon layer and a 2 μm buried oxide (BOX) layer on a 675 μm thick silicon substrate. Two π-PSs are strategically embedded at $1/3\ L$ and $2/3\ L$ positions along the 4PS-SBG cavity. As depicted in Figure 6(b), the side-wall gratings of the 4PS-SBG are symmetrically arranged, featuring a seed grating period ($\Lambda_0$) of 351 nm, a grating recess ($d$) of 25 nm, and a ridge waveguide width ($W$) of 520 nm. The chirped 4PS-SBG waveguide has an effective refractive index of 2.47 at the wavelength of 1535 nm. The grating samplings adhere to equation (5). The device comprises 96 complete sampling units distributed throughout its cavity, denoted by $n = 96$. The chirp rate $C$ is set at 200 nm/mm, with 1st sampling period $P_1$ of 3.083 μm, aligning the +1st order channel near the wavelength of 1530 nm. The 60 nm difference in sampling periods between $P_1$ and $P_n$ can be precisely achieved using electron-beam lithography (EBL), which offers a resolution of 0.5 nm.

The fabrication process of the device begins with defining the 210-nanometer-high silicon ridge waveguides and sidewall gratings. This requires the use of hydrogen silsesquioxane (HSQ) resist mask, EBL, and inductively coupled plasma (ICP) etching. During etching, an SPTS Rapier DSiE system is used with a gas flow of $C_4F_8/SF_6$ at a ratio of 90:30 sccm. Afterward, the HSQ mask is removed using a hydrofluoric acid (HF) solution, leaving a 10 nm residual silicon layer to protect the underlying silicon dioxide. A second round of EBL exposure and dry etching is then performed using polymethyl methacrylate (PMMA, AR-P 642 200 k Anisole 12%, baked at 180°C, and developed in a 2.5:1 Isopropyl Alcohol (IPA) solution) as resist and masks. This process forms 110 nm deep GC at both the input and output ends of the 4PS-SBG. Next, to create the top cladding and prepare for subsequent metal deposition, $SiO_2$ deposition and HSQ spin-coating are used to planarize the surface. A 400 nm thick $SiO_2$ layer is deposited on the wafer surface using plasma-enhanced chemical vapor deposition (PECVD), forming the embedded waveguide. Afterward, a 600-nm-thick HSQ layer is spin-coated and annealed at 180°C. Measurements using a Bruker Dektak XT Stylus Profiler showed that the height difference at the waveguide section was reduced from 210 nm to 50 nm, confirming successful surface planarization. Finally, a two-layer PMMA resist (AR-P 642 200k Anisole 15% PMMA and AR-P 679 950k Ethyl Lactate 2% PMMA, baked at 180°C, developed in a 2.5:1 IPA solution) was used in combination with EBL and a lift-off technique. Metal evaporation was then employed to define the MH components, consisting of 20 nm Ti, 160 nm Pt, and 50 nm Au.

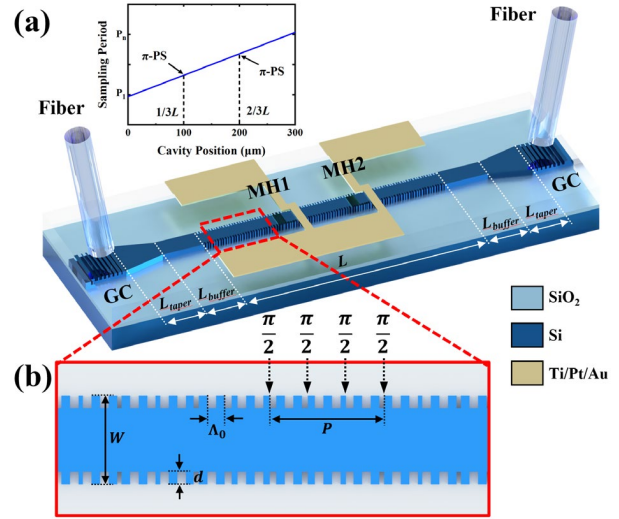

**Figure 6.** (a) Schematic representation of the photonic filter with grating coupler, 4PS-SBG and MHs. (b) Enlarged schematic of the 4PS-SBG.

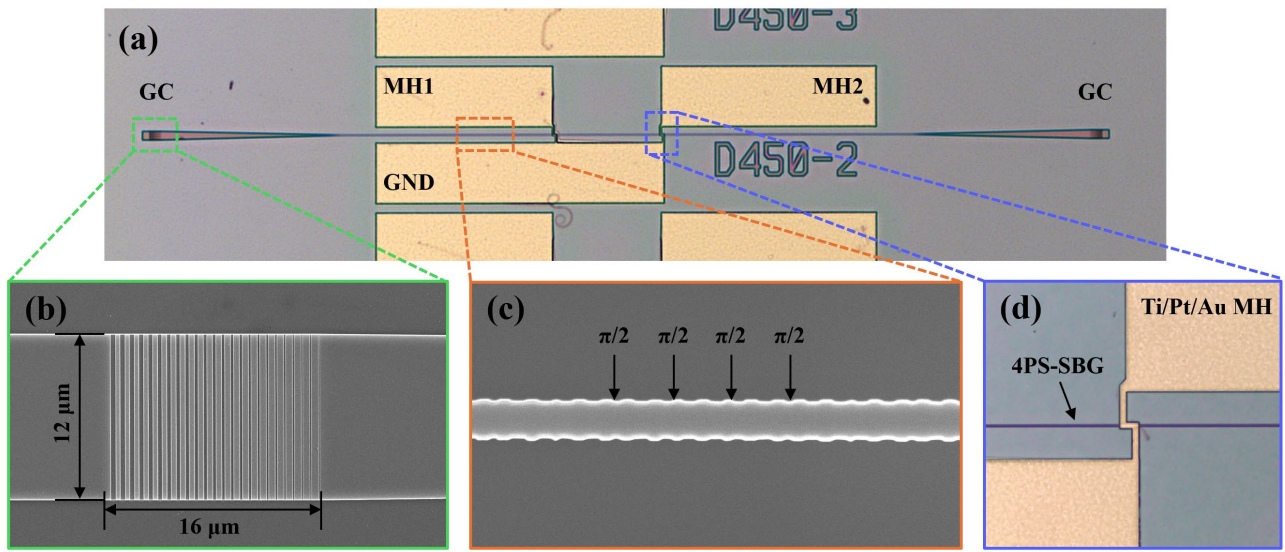

**Figure 7.** (a) Optical microscope images of the device after fabrication, and SEM images of the grating coupler (b), SEM image of the 4PS-SBG (c), and optical microscope image of the resistor wire of the MH (d).

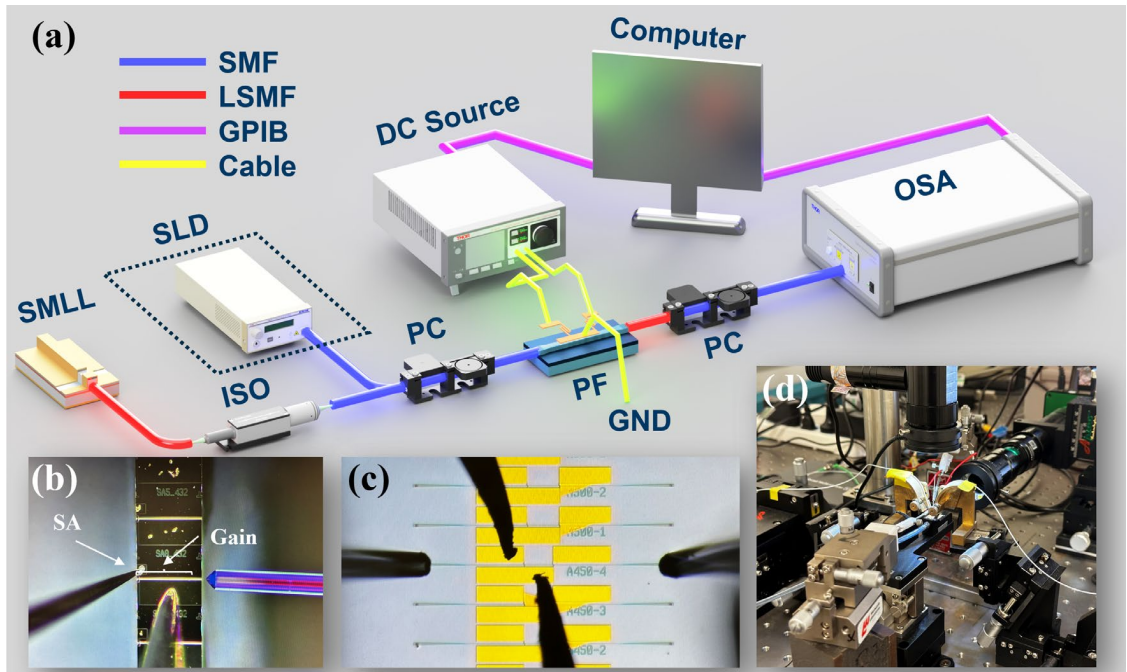

**Figure 8**. (a) Schematic diagram of characterization setup, (b) light source coupled from the SMLL using a lensed fiber, (c) Microscope image of the PF monolithically integrated with GCs, showing metal probes connected to MHs and cleaved SMFs connected to the GCs, (d) photograph of measurement stage.

This surface planarization method is designed to simplify the fabrication process and optimize yield. In typical MH circuits, delamination of the resistive layer is a common primary failure mechanism. This issue arises from excessive interfacial stress caused by a combination of surface roughness and high temperatures. Once the resistive layer delaminates, localized overheating occurs, which further accelerates the delamination process, eventually leading to the burnout of the resistive wire.

In the fabrication process, we systematically optimized the photoresist thickness, electron beam dose, and beam step size (BSS) in the EBL, guided by a series of controlled dose-response and fabrication tests. This careful optimization was crucial for achieving high-resolution sub-wavelength grating periods, smooth sidewall profiles, and precise control over the grating recess.

The top-view optical microscope image of the fabricated filter device is depicted in Figure 7(a). Figure 7(b) presents a zoomed-in SEM image of the fabricated GC, while Figure 7(c) showcases the fabricated 4PS-SBG structure. Additionally, Figure 7(d) illustrates the MHs fabricated on the surface of the cladding and its relative position to the embedded grating waveguide. The MH consists of contact pads and heating resistance wires, with serpentine heating wires positioned directly above the two PSs regions, each with a width of 2 μm, and a resistance of 8.9 Ω at 20°C.

In Figure 8(a), the experimental setup for device characterization at room temperature includes a superluminescent diode (SLD) as the light source, with a center wavelength of 1551 nm and a 3 dB bandwidth of 30 nm, as well as a 100 GHz frequency spacing SMLL with a center wavelength of 1535 nm. A single-mode fiber (SMF) is connected to the SLD, and a polarization controller directs the light toward the input GC surface. At the output end, a lensed single-mode fiber (LSMF) with coating is used to couple the output light from the GC surface, minimizing unwanted resonances from a Fabry-Pérot (FP) cavity formed between the fiber and the GC. The input and output SMFs are positioned directly above the GCs, with a vertical offset angle of 15°. The output optical signal is observed using an optical spectrum analyzer (OSA) with a resolution bandwidth (RBW) of 0.06 nm. A polarization controller (PC) is placed before the OSA to ensure that only the TE mode light is characterized. The output wavelength of the device can be adjusted by varying the injection current applied to the MH contact pads ($I_{MH}$).

It is worth noting that during the continuous tuning experiments of the filter, only the SLD was used as the light source. For subsequent optical frequency division applications, a 100 GHz SMLL is used as the input source (see Figure 8(b)) When using the SMLD as the input source, an optical isolator (ISO) is also introduced to prevent reflected light from the PF from damaging the SMLL.

To demonstrate the filtering properties and associated parameters, we applied varying currents to the MH pads and measured the resulting resonant peak shifts in the transmission spectrum, as shown in Figure 8(c). An automated measurement system was designed and implemented, using a general-purpose interface bus (GPIB) to interface seamlessly with the measurement instruments. Controlled via LABVIEW software, the system enabled efficient and rapid data acquisition. This setup not only facilitates high-throughput measurements but also ensures consistent precision across multiple experimental runs. Figure 8(d) shows a photograph of the measurement stage in our test lab.

## RESULTS AND DISCUSSION

**Integrated Photonics Filter Characterization.** Figure 9(a) shows the transmission spectrum when only $I_{MH1}$ is tuned from 0

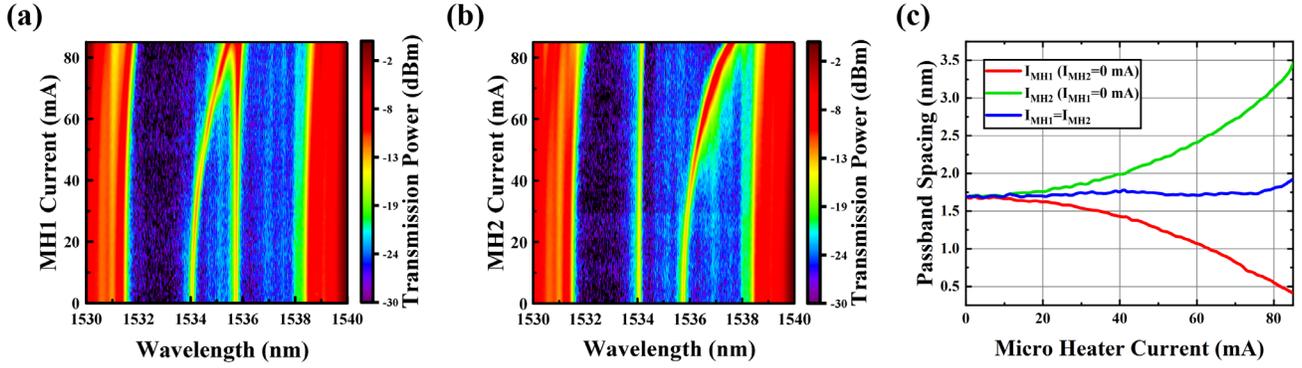

**Figure 9.** (a) 2D optical spectra with only MH1 modulation; (b) 2D optical spectra with only MH2 modulation; (c) wavelength gap of the two passbands versus MH currents.

mA to 85 mA. In which when the modulation current is 0 mA, the spectrum reveals dual passbands (Q factor: $1.18 \times 10^4$) located within a 7.5 nm range, centered at the 1534.9 nm stopband. The frequency separation is 216.3 GHz, closely matching the simulation results and demonstrating a balanced and consistent intensity distribution. The 4.1 nm redshift of the center wavelength compared to the simulation result may be attributed to etching the ridge waveguide to a height of 210 nm instead of 220 nm as in the simulation. This reduction in height increases the effective index, which in turn shifts the center wavelength. The relatively lower Q factor compared to the simulation may be due to the inadequate RBW of OSA and fabrication imperfections, such as sidewall and surface roughness of the ridge waveguide, which introduce additional scattering losses and thus reduce the Q factor. The effective grating coupling coefficient $\kappa$ for the +1st channel is measured to be 241/cm and measured propagation loss of the designed 4PS-SBG waveguide is 10.9 dB/cm at a wavelength of 1550 nm.

Figure 9(a) also illustrates that during the scan of $I_{MH1}$ from 0 to 85 mA, the thermo-optic effect of the PS induces a redshift of passband 1. The center wavelength of passband 2 remains unchanged during MH1 modulation, allowing for independent tuning of a single band. The separation frequency changes from 216.3 GHz to 52.1 GHz. The consistent positioning of the stopband suggests precise alignment and geometric layout of the MHs, with minimal residual heat leakage. Similarly, as depicted in Figure 9(b), independent modulation of passband 2 can also be achieved by solely tuning $I_{MH2}$. When $I_{MH2}$ is tuned from 0 mA to 85 mA and $I_{MH1}=0$ mA, the separation frequency changes from 216.3 GHz to 439.5 GHz. Furthermore, as shown in Figure 9(c), analysis of the changes in passband spacing during MH modulation from 0 to 85 mA reveals a continuous modulation range from 52.1 GHz to 439.5 GHz, enabling dynamic tuning of passband separation. According to the simulation results in Figures 4(a) and 4(b), the tuning wavelength gap can up to 661 GHz if the MH wire is optimized to prevent breakage when a high injection current (>85 mA) is applied. Experiments on synchronous modulation of the dual passbands by tuning $I_{MH1}$ and $I_{MH2}$ from 0 mA to 85 mA simultaneously were also conducted, maintaining a relatively stable wavelength spacing and further confirming the independent tuning characteristic of the dual passbands.

**Optical Frequency Division Experiments.** For the optical frequency division experiments, a passive SMLL is designed and fabricated using the AlGaInAs/InP material system to produce an optical frequency comb (OFC). It features an asymmetric multiple-quantum-well (MQW) epilayer structure. Figure 8(b) shows an optical microscope picture of the 100 GHz repetition frequency SMLL device. The length of the entire cavity length is 432 μm. The length of the saturable absorption (SA) section is 10 μm, the gain section is 412 μm, and the isolation groove between the gain and SA sections is 10 μm. Its ridge waveguide is 2.5 μm wide. Passive mode locking (ML) is achieved by forward biasing the gain section and applying a reverse voltage to the saturable absorber (SA) section. To more accurately characterize the optical frequency division performance of the PF, experiments were conducted with the SMLL set to a central wavelength of 1535 nm under pure ML conditions. This was done with a gain current of 100 mA and a reverse bias voltage of -2.3 V on the SA section. The SMLL exhibited a full width at half maximum (FWHM) of 4.5 nm, generating an OFC with a 100 GHz spacing, as shown in Figure 10. The output optical power from the gain facet was 10 mW. The isolated laser optical signal, after passing through the ISO, was coupled into the PF. A direct current (DC) electrical signal from a current source controller was applied to the MH via contact pads. The optical signal output from the PF was then transmitted to an OSA after the polarization direction was adjusted by the PC. The output spectra shown in Figures 11(a) to 11(d) were obtained by modulating the driving currents of MH1 and MH2, as depicted in the respective figures. The blue solid line represents the output optical spectrum after the PF, while the pink dashed line indicates the reference output spectrum of the SMLL after passing through a ridge waveguide of the same length as the filter. An analysis of the transmission characteristics of the dual wavelength PF revealed that the output spectra closely followed the trends observed using the SLD source, as shown in Figure 9(c). Specifically, when the current applied to MH1 was increased from a low to high value, passband 1 shifted towards passband 2, reducing the separation frequency of the PF. In contrast, modulation of MH2 caused passband 2 to shift away from passband 1, increasing the separation frequency. Figures 11(a) to 11(d) illustrate four distinct filtering configurations with separation frequencies of 100 GHz, 200 GHz, 300 GHz, and 400 GHz, demonstrating a tunable frequency division range from 100 GHz to 400 GHz. These configurations correspond to the modulation currents $I_{MH1}$ and $I_{MH2}$ shown in the figures. When tuning the PF to select the 200 GHz optical frequency spacing (Figure 11(b)), the maximum SMSR exceeds 10 dB. The measured minimum insertion loss at peak (1536.7 nm) transmission is 0.85 dB.

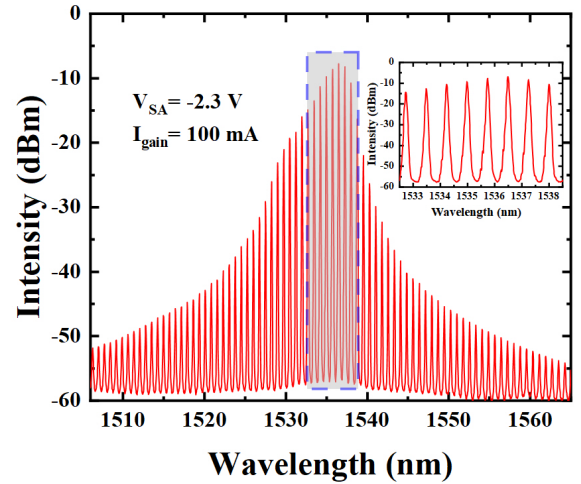

**Figure 10.** Optical spectrum of the gain section facet of the SMLL in the ML state with a gain section injection current of 100 mA and a reverse bias voltage of -2.3 V applied to the SA section.

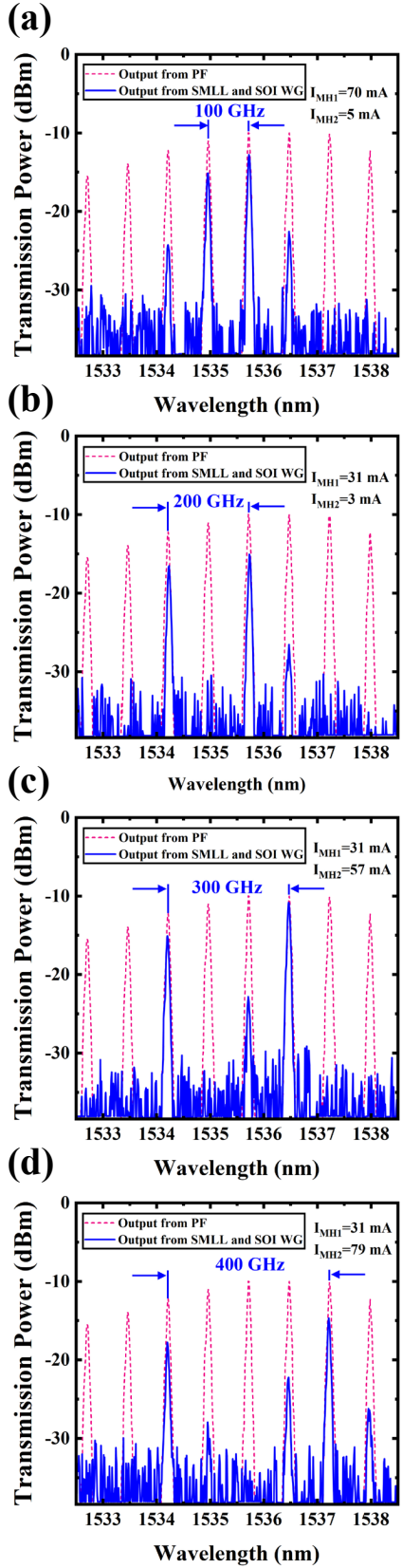

**Figure 11.** Output spectrum of the OFC signal from the SMLL and SOI ridge waveguide (dashed line), and the OFC filtered by the PF with spacings of (a) 100 GHz, (b) 200 GHz, (c) 300 GHz, and (d) 400 GHz (solid line).

## CONCLUSIONS

In summary, we have proposed a dual-frequency independently tunable PF integrated on a SOI platform. Both simulation and experimental results demonstrate that the introduction of equivalent chirp enables independent photon distribution in the dual passbands with minimal thermal crosstalk from the metal heaters (MHs), allowing for independent modulation of the passbands. The adoption of the 4PS-SBG structure enhances the extinction ratio and effectively suppresses unwanted side modes. The filter exhibits continuous frequency tuning capabilities from 52.1 GHz to 439.5 GHz, with an unprecedentedly wide tuning range of 387 GHz and a narrow passband featuring a -3 dB spectral resolution of 16 GHz. Using a SMLL with 100 GHz frequency spacing demonstrated the PF's optical frequency division capability from 100 GHz to 400 GHz, achieving a SMSR exceeding 10 dB. Our work addresses the challenges of miniaturization and wideband tunability in multi-frequency filters on a silicon photonic platform. This technology can also be extended to terahertz (THz) PF applications. With further optimization of the device and integration of lasers, a tunable fully integrated microwave photonics (MWP) system based on grating filters could be realized. This advancement paves the way for a new class of integrated platforms capable of performing a range of on-chip MWP processing functions, including phase shifters, frequency converters, and RF sources.

## AUTHOR INFORMATION

**Corresponding Author**

Simeng Zhu - Email: 2635935z@student.gla.ac.uk.

**Author Contributions**

The manuscript was written through contributions of all authors. All authors have given approval to the final version of the manuscript.

**Funding**

This work was supported by the U.K. Engineering and Physical Sciences Research Council (EP/R042578/1).

**Notes**

The authors declare no competing financial interests.

## ACKNOWLEDGMENT

We would like to acknowledge the staff of the James Watt Nanofabrication Centre at the University of Glasgow for their help in fabricating the devices.